\documentclass[smallextended]{svjour3}       % onecolumn (second format)
\smartqed  % flush right qed marks, e.g. at end of proof
\usepackage{graphicx}
\begin{document}

\title{Reconstruction of $f(R)$ Lagrangian from a massive scalar field}

\author{Soumya Chakrabarti         \and
        Jackson Levi Said         \and
        Kazuharu Bamba
}

\institute{Soumya Chakrabarti \at
              Centre for Theoretical Studies,\\
              Indian Institute of Technology, Kharagpur,\\
              West Bengal 721 302, India.
              \email{soumya@cts.iitkgp.ernet.in}           %  \\
           \and
           Jackson Levi Said \at
              Institute of Space Sciences and Astronomy,\\
              University of Malta, Msida,\\
              MSD 2080, Malta.
              Department of Physics,\\
              University of Malta, Msida,\\
              MSD 2080, Malta.
              \email{jackson.said@um.edu.mt}
           \and
           Kazuharu Bamba \at
              Division of Human Support System,\\
              Faculty of Symbiotic Systems Science,\\
              Fukushima University,\\
              Fukushima 960-1296, Japan.
              \email{bamba@sss.fukushima-u.ac.jp}
}

\date{Received: date / Accepted: date}

\maketitle

\begin{abstract}
A reconstruction scheme from a minimally coupled self-interacting scalar field is discussed in the regime of $f(R)$ gravity. This involves a direct way to solve the system of equations for some well-known form of self-interaction potential. For a power law or combination of power law potential, the scheme originates from a theorem of invertible point transformation and integrability of anharmonic oscillator equations. A case with exponential self-interaction potential is also included. The viability of the $f(R)$ models is discussed in brief.
\end{abstract}

\section{Introduction}
More than a century has passed since general relativity (GR) was formulated by Albert Einstein. The field equations of the theory work as the defining relation between the curvature of spacetime and the energy-momentum of the system under consideration. Solutions of the field equations govern different aspects of gravitational physics, ranging over a vast scale of phenomena. For instance, under the assumption of homogeneity and isotropy the field equations produce the famous Friedmann-Lema\^{i}tre-Robertson-Walker (FLRW) metric \cite{Weinberg:1972kfs} of cosmology. However, recent developments of observational cosmology suggests that the universe has seen non-trivial phases of cosmic acceleration. The phase of an early inflation is necessary to resolve the horizon and flatness problems of GR \cite{Linde:2007fr,Guth:1980zm}. Apart from this, a late-time accelerating phase \cite{Riess:1998cb,Perlmutter:1998np} exists, thought to be an artefact of the dark energy component, a fluid with effective negative pressure. For comprehensive reviews we refer to the works of Clifton, Ferreira, Padilla and Skordis \cite{Clifton:2011jh}, Nojiri, Odintsov and Oikinomou \cite{Nojiri:2017ncd}, Koyama \cite{Koyama:2015vza}. \\

To account for the late-time acceleration, perhaps the simplest remedy is to use a cosmological constant in the GR field equations. Wang, Caldwell, Ostriker and Steinhardt \cite{Wang:1999fa} discussed the observational aspects of cosmological models including matter alongwith a special case of cosmological constant. A cosmological constant can in principle reproduce almost all of the cosmological observations, extensively discussed by Spergel et al. \cite{sperg}, Eisenstein et al. \cite{eisen}, Jain and Taylor \cite{jain}. However, this avenue has a fine-tuning problem and the energy scale predicted for the vaccuum is severely mismatched with that found from a quantum theory, discussed by Weinberg \cite{Weinberg:1988cp}. Another option of modifying the energy-momentum description of gravity is to introduce a scalar field with a slowly varying potential. This approach has served successfully as a candidate for both early inflation and the late time acceleration. For a very recent summary we refer to the monograph of Bahamonde et al.\cite{Bahamonde:2017ize}. Without considering additional scalar fields, one may also resort to some modification of the Einstein-Hilbert action, which may resolve some or all of the consistency issues while also preserving the healthy avenues of GR, for instance, the solar system tests. \medskip

The most straightforward and popular modification comes from gravitational actions non-linear in the Ricci scalar $R$, namely, the $f(R)$ theories of gravity (for extensive reviews see for instance the works of Sotiriou and Faraoni \cite{soti1}, Carroll, Duvvuri, Trodden and Turner \cite{carro}, Kerner \cite{kerne}, Teyssandier \cite{teyss}, Magnano, Ferraris and Francaviglia \cite{magna}. The models first became popular in the $1980$-s since Nojiri and Odintsov proved \cite{Nojiri:2003rz} that they can be derived from fundamental physical theories, for example the M-theory. They naturally admit a phase of accelerated expansion driven by geometry which can easily be associated with inflation, discussed by Starobinsky \cite{staro}, Stelle \cite{stelle}. Similarly, dark energy could be thought of having a geometrical origin rather than adding a vacuum energy or an additional scalar field to the energy-momentum tensor at the outset. For more recent reviews on modified gravity theories and dark energy problem, we refer to the works of De Felice and Tsujikawa \cite{DeFelice:2010aj}, Nojiri and Odintsov \cite{Nojiri:2010wj}, Capozziello and De Laurentis \cite{Capozziello:2011et}, Faraoni and Capozziello \cite{Capozziello:2010zz}, Bamba and Odintsov \cite{Bamba:2015uma}, Bamba, Capozziello, Nojiri and Odintsov \cite{Bamba:2012cp}. \medskip

Number of cosmological solutions in these theories are rather limited, owing to the nonlinearity of the field equations (fourth order in metric components). The difficulty can be reduced somewhat by using the theory of dynamical systems, which provides a relatively simple method for obtaining a qualitative description of the global dynamics of these models for a given $f(R)$. Usefulness of a dynamical system analysis is extensively studied over the years for different cosmological setups \cite{carloni}. Another approach is to assume that the evolution of the universe is known, and to invert the field equations to find out what class of $f(R)$ theories allow the aforementioned evolution. Popularly known as a reconstruction, this approach has received considerable attention lately, for example in the works of Nojiri and Odintsov \cite{Nojiri:2006gh,Nojiri:2006be}. They developed a method for $f(R)$ reconstruction from a proper cosmological dynamics, compatible with different viability criterion such as the solar system tests. They further extended the reconstruction scheme for a number of other modifications of gravity, for instance, the scalar-tensor theory, $f(G)$, scalar-Gauss-Bonnet theories. Similar methods of reconstruction are very popular and thereafter a plethora of reconstruction models have been proposed in literature \cite{nojiri}. In a very recent work on cosmological reconstruction, Goheer, Larena and Dunsby \cite{goheer} proved that an exact power-law scale factor and a perfect fluid cosmology is only allowed for a power law $f(R)$ theory. Goswami, Odintsov, Dunsby and Elizalde \cite{goswami} studied cosmological reconstruction in $f(R)$ theories, quite recently. More extensive analysis of reconstruction methods has been carried out by Carloni, Goswami and Dunsby \cite{cgd}, Carloni, Dunsby, Capozziello and Troisi \cite{Carloni:2004kp}, Nojiri and Odintsov \cite{Nojiri:2006ri}, Bamba, Myrzakulov, Nojiri and Odintsov \cite{Bamba:2012vg}, where specific $f(R)$ theories have been treated giving a smooth deceleration-acceleration transition in cosmology.  \medskip

In the present work we give a strategy that can lead one to the reconstruction of modifed gravity Lagrangians from a self-interacting scalar field as the matter distribution. The scalar field is minimally coupled in the action. The starting motivation of this work is mathematical. Apparently, one does not need the assumption of a known expansion history of the Universe at the outset. The treatment originates from  identifying the scalar field evolution equation as an anharmonic oscillator equation. We assume that under a set of invertible point transformations, the evolution equation can be point transformed into an integrable form. The criterion for validity of such a transformation can be written in the form of a theorem of integrability. We use the integrability criterion to solve for the scale factor and the scalar field from just one equation. Then we use the solutions to reconstruct and write explicitly the functional form of $f(R)$ that allows such a cosmological dynamics, for different choices of the self-interaction potential, using the other field equations. In a similar manner, different aspects of gravitational physics involving the collapsing dynamics of a massive scalar field is discussed at length by Chakrabarti and Banerjee \cite{scnb2,Banerjee:2017njk}. A similar setup was studied under the regime of Scalar-Einstein-Gauss-Bonnet gravity by Chakrabarti \cite{Chakrabarti:2017apq}). \\

The organization of this manuscript is as follows. Section $II$ deals with the basic action and field equations under consideration. In section $III$, we consider a simple power-law self-interaction potential of the scalar field and the dynamics of cosmological scale factor and the reconstruction of $f(R)$. Sections $IV$ and $V$ deal with a combination of power law potentials and an exponential potential. We discuss the viability of $f(R)$ models in section $VI$ and conclude the manuscript in section $VII$.

\section{Action and the Field Equations}
We consider the minimally coupled scalar field action for $f(R)$ gravity as
\begin{equation}\label{action}
S = \frac{1}{2\kappa^2}\int \sqrt{-g}d^4x \{ f(R) - \phi_{,\mu}\phi_{,\nu}g^{\mu\nu}-2V(\phi) \}\,.
\end{equation}
$\phi$ is the self-interacting scalar field with the energy-momentum tensor described by
\begin{equation}\label{SET}
T_{\mu\nu} = \phi_{,\mu}\phi_{,\nu} - g_{\mu\nu}\left(\frac{1}{2}\phi^{,\alpha}\phi_{,\alpha} + V(\phi)\right)\,,
\end{equation}
where $V(\phi)$ is the potential, and $\kappa^2$ is set equal to unity so that geometric units are considered throughout. The first term in Eq. (\ref{action}) represents the generalization of the Einstein-Hilbert action, while the other two terms represent kinetic and potential energies of the scalar field. \\

\noindent In this work, we consider a spatially flat Universe with FLRW metric
\begin{equation}\label{frw}
ds^2 = -dt^2+a^2(t)( dr^2 + r^2( d\theta^2 + \sin^2\theta \; d\varphi^2) )\,,
\end{equation}
where $a(t)$ is the cosmological scale factor. The field equations can then be written as
\begin{eqnarray}\label{fe1}
H^2&=\frac{1}{3F}(K+V) +\frac{R}{6}-\frac{f}{6F}-H\frac{\dot{F}}{F}\,,\\
\dot{H} + H^2&= -\frac{1}{3F}(2K-V) -\frac{f}{6F}+\frac{R}{6}-\frac{H\dot{F}}{2F}-\frac{\ddot{F}}{2F}\,,
\label{fe2}
\end{eqnarray}
where $F(R)=f'(R)=\frac{df(R)}{dR}$ and $K=\frac{1}{2} \dot{\phi}^2$ is the kinetic energy of the scalar field $\phi$. Overdots denotes differentiation with respect to cosmic time and prime denotes differentiation with respect to the scalar curvature. Varying the action Eq.(\ref{action}) with respect to $\phi$ leads to the scalar field equation results in

\begin{equation}\label{scalarkg}
\ddot\phi + 3 \frac{\dot{a}}{a}\dot\phi + \frac{dV}{d\phi} = 0\,.
\end{equation}

The system of equations governing the dynamics is therefore defined by Eq.(\ref{fe1}) and Eq.(\ref{scalarkg}). Our first aim is to integrate the scalar field evolution equation Eq.(\ref{scalarkg}) straightaway. The criterion for such an integration is defined in terms of an invertible point transformation, worked out by Duarte, Moreira, Euler and Steeb \cite{duarte}, Euler, Steeb and Cyrus \cite{euler}, Euler \cite{euler1}, Harko, Lobo and Mak \cite{harko}. An anharmonic oscillator takes the form of a nonlinear second order differential equation with variable coefficients as
\begin{equation}
\label{gen}
\ddot{\phi}+f_1(t)\dot{\phi}+ f_2(t)\phi+f_3(t)\phi^n=0\,,
\end{equation}
where $f_i$ are functions of $t$ and $n$ is a constant ($n \in {\cal Q}$). This equation can be integrated in a straightforward manner under certain conditions and the essence of the integrability criterion is that, an equation of the form Eq.(\ref{gen}) can be point transformed into an integrable form. The necessary and sufficient condition for such a transformation is that the exponent $n\notin \left\{-3,-1,0,1\right\} $ and the coefficients of Eq.(\ref{gen}) satisfy the differential condition

\begin{eqnarray}\nonumber
\label{int-gen}
&&\frac{1}{(n+3)}\frac{1}{f_{3}(t)}\frac{d^{2}f_{3}}{dt^{2}} - \frac{(n+4)}{\left( n+3\right) ^{2}}\left[ \frac{1}{f_{3}(t)}\frac{df_{3}}{dt}\right] ^{2} + \frac{(n-1)}{\left( n+3\right) ^{2}}\left[ \frac{1}{f_{3}(t)}\frac{df_{3}}{dt}\right] f_{1}\left( t\right) \\&& 
+ \frac{2}{(n+3)}\frac{df_{1}}{dt} + \frac{2\left( n+1\right) }{\left( n+3\right) ^{2}}f_{1}^{2}\left( t\right) = f_{2}(t)\,.
\end{eqnarray}

\noindent Thereafter one can then introduce a pair of new variables $\Phi$ and $T$ given by 
\begin{eqnarray}
\label{Phi}
\Phi\left( T\right) &=&C\phi\left( t\right) f_{3}^{\frac{1}{n+3}}\left( t\right)
e^{\frac{2}{n+3}\int^{t}f_{1}\left( x \right) dx }\,,\\
\label{T}
T\left( \phi,t\right) &=&C^{\frac{1-n}{2}}\int^{t}f_{3}^{\frac{2}{n+3}}\left(
\xi \right) e^{\left( \frac{1-n}{n+3}\right) \int^{\xi }f_{1}\left( x
\right) dx }d\xi \,,\nonumber\\
\end{eqnarray}%
where $C$ is a constant. Eq.(\ref{gen}) can then be transformed into  
\begin{equation}
\label{Phi1}
\frac{d^{2}\Phi}{dT^{2}}+\Phi^{n}\left( T\right) = 0\,.
\end{equation}

One can write the scalar field $\phi$ as a function of $t$ by putting back the transformations
\begin{eqnarray}\nonumber
\label{phigen}
&&\phi\left( t\right) = \phi_{0}\Big[ C^{\frac{1-n}{2}}\int^{t}f_{3}^{\frac{2}{n+3}}\left( \xi \right) e^{\left( \frac{1-n}{n+3}\right) \int^{\xi }f_{1}\left(x \right) dx }d\xi - T_{0}\Big]^{\frac{2}{1-n}}\\&&  f_{3}^{-\frac{1}{n+3}}\left( t\right) e^{-\frac{2}{n+3}\int^{t}f_{1}\left( x \right) dx}\,,
\end{eqnarray}
where $\phi_{0}$ and $T_0$ are nonvanishing constants of integration and $C$ is born out from the definition of the point transformations (\ref{Phi}) and (\ref{T}).

\section{Power Law Self-Interaction}
In this section, we assume the self-interaction of the scalar field to be proportional to a power law function of the scalar field or $\frac{dV(\phi)}{d\phi} \sim \phi^{n}$. The effective mass of the field is therefore given by $\frac{d^{2}V(\phi)}{d\phi^{2}}$ at $\phi = 0$. Such forms of self-interaction potentials have played very importnat roles in scalar field cosmology, for example, inverse powers of the field work as brilliant fit as they demonstrate the tracker quintessance behavior, discussed by Ratra and Peebles \cite{Peebles:2002gy} for a potential $V(\phi) \sim \frac{M^{(4+\alpha)}}{\phi^{\alpha}}$ ($M$ being the Planck mass). We also refer to the works of Zlatev, Wang and Steinhardt in this regard \cite{Zlatev:1998tr,Steinhardt:1999nw}. \\

For a simple power law potential, a comparison with the anharmonic oscillator equation gives one the coefficients in the form $f_{1} = 3\frac{\dot{a}}{a}$, $f_{2} = 0$ and $f_{3} = 1$. Therefore the criterion of integrability as in Eq.(\ref{int-gen}) gives a differential equation governing the time evoltion of the scale factor as

\begin{equation}
\dot{a}^2 a^{\frac{4n}{(n+3)}} = \lambda^2\,.
\end{equation}

\noindent This can be solved to find the allowed time evolution of the scale factor as

\begin{equation}\label{exactscale1}
a(t) = [\lambda (t-t_{0})]^{\frac{(n+3)}{3(n+1)}}\,,
\end{equation}
where it is assumed that $\dot{a} > 0$.
\begin{figure}[h]
\begin{center}
\includegraphics[scale=0.7]{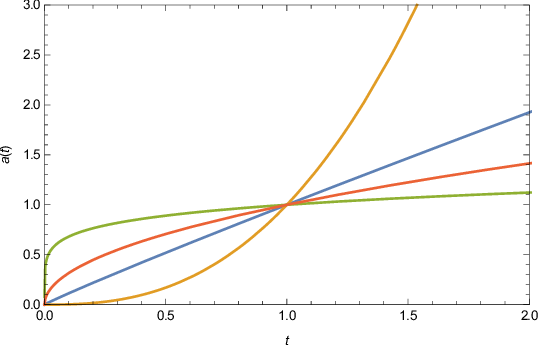}
\caption{Evolution of the scale factor for $V(\phi) \sim \phi^{(n+1)}$. The code of color for different graphs are defined as; $n$ -- Red: $n = 3$, Blue: $n = 0.09$, Yellow: $n = -0.7$, and Green: $n = -5$.}
\label{power_law_graph}
\end{center}
\end{figure}

Fig. \ref{power_law_graph} represents the evolution of the scale factor with time for different values of $n$. It is evident that, the nature of the exponent governs the evolution, for instance, for $n > 0$ one has a time evolution similar to flat cosmology. On the other hand, one has an evolution mimicking open cosmologies with marginally negative values of $n$. In the region where $-1 < n < 0$ the time evolution hints at a possibility of a late-time acceleration. \medskip

\begin{figure}[h]
\begin{center}
\includegraphics[scale=0.7]{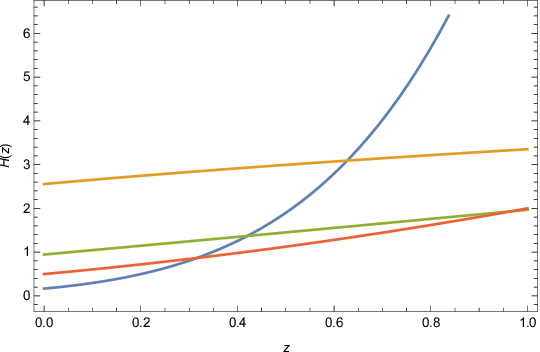}
\caption{Hubble parameter as a function of $z$ for the power-law potential. The code of color for the index $n$ are the same as in Fig.(\ref{power_law_graph})}
\label{Hubble_powerlaw}
\end{center}
\end{figure}

With an exact form of the scale factor as a function of time, one can calculate the functional form of the scalar field from Eq.(\ref{phigen}) as 
\begin{equation}\label{exactfield}
\phi(t) = \xi_{0} (t-t_{0})^{\frac{2}{(1-n)}}\,.
\end{equation}
$\xi_{0}$ is a constant. The scalar field maintains a positive profile as long as $\xi_{0} > 0$. Evolution of the Hubble parameter as a function of the redshift shows a steadily increasing behavior as a function of $z$ for all choices of $n$, as shown in Fig. \ref{Hubble_powerlaw}. 

\subsection{Reconstruction of $f(R)$}
We re-write the exact time evolution of the scalar field and the scalar field as $a(t) = \lambda (t-t_{0})^{\lambda_1}$ and $\phi(t) = \lambda_{2} (t-t_{0})^{\lambda_3}$ respectively. The aim is to study the Friedmann equation in Eq.(\ref{fe1}) in order to solve for an exact form of $f(R)$ as a function of $R$. 
The Friedmann equation takes the form

\begin{equation}\label{fee}
F H^2 = \frac{1}{3} \Big(\frac{\dot{\phi}^2}{2} + V(\phi) \Big) +\frac{F R}{6} - \frac{f}{6} - H \dot{F}\,.
\end{equation}

From the definition of the Ricci scalar $R = 6(\dot{H}+2H^2)$, one can obtain time as a function of the Ricci scalar, and write
\begin{equation}
(t-t_{0}) = [6 \lambda_{1} (2 \lambda_{1} - 1)]^{\frac{1}{2}} R^{-\frac{1}{2}}\,.
\end{equation}

\noindent Without any loss of generality one may choose $\lambda = 1$ and 
\begin{eqnarray}\nonumber
\lambda_{1} = \frac{(n+3)}{3(n+1)}\,, \nonumber\\
\lambda_{3} = \frac{2}{(1-n)}\,.
\end{eqnarray}

$\lambda_2$ can be evaluated from the consistency of the scalar field evolution equation as 

\begin{equation}
\lambda_2^{2} = \Bigg[\frac{2(n+3)}{(n^2 -1)} - \frac{2(n+1)}{(n-1)^2}\Bigg]^{\frac{2}{(n-1)}}\,.
\end{equation}
Using the definition of the scalar field and the Ricci scalar one can then write the scalar field equation as

\begin{equation}
\frac{\dot{\phi}^{2}}{2} + \frac{\phi^{(n+1)}}{(n+1)} = \lambda_{4} R^{\lambda_{5}}\,.
\end{equation}

\noindent The coefficients $\lambda_{4}$ and $\lambda_{5}$ are defined by the relations

\begin{eqnarray}\nonumber\nonumber
& \lambda_{4} = \Big[\frac{2\lambda_{2}^2}{(1 - n)^2} + \frac{\lambda_{2}^{(n+1)}}{(n+1)}\Big]\Big[6\lambda_{1}(2\lambda_{1} - 1)\Big]^{\frac{(1+n)}{(1-n)}}\,, \nonumber\\
& \lambda_{5} = \frac{(n+1)}{(n-1)}\,.
\end{eqnarray}

\noindent Moreover, we write the Hubble parameter $H = H(R)$ as
\begin{equation}
H = \lambda_{1} [6\lambda_{1}(2\lambda_{1}-1)]^{-\frac{1}{2}} R^{\frac{1}{2}}\,.
\end{equation}

Now we simply change the derivatives with respect to cosmic time $t$ into derivatives with respect to the Ricci scalar $R$ and write the Friedmann equation in the new form as

\begin{eqnarray}\nonumber
\label{Mequation1}
&&\lambda_{1}^{2} [6\lambda_{1}(2\lambda_{1} - 1)]^{-1} R\frac{df(R)}{dR} = \frac{1}{3} \lambda_{4} R^{\lambda_5} + \frac{R}{6} \frac{df(R)}{dR} - \frac{f(R)}{6}\\&&
- \lambda_{10} R^2 \frac{d^{2}f(R)}{dR^2}\,,
\end{eqnarray}
where
\begin{equation}
\lambda_{10} = 12\lambda_{1}^{2} (1 - 2\lambda_{1}) [6\lambda_{1}(2\lambda_{1} - 1)]^{-2}\,.
\end{equation}

A solution of Eq.(\ref{Mequation1}) gives one the general functional form of $f(R)$ as a function of $R$ as we write below

\begin{equation}\label{solution}
f(R) = A_{0} R^{\lambda_{5}} + A_{1} R^{(\lambda_{19} + \lambda_{13})} + C_{1} R^{\lambda_{25}} + C_{2} R^{\lambda_{26}}\,,
\end{equation}
where
\begin{eqnarray} \nonumber
&& A_{0} = B_{0} [B_{1} - B_{2}]\,, \\&& \nonumber
A_{1} = B_{0} [B_{1} + B_{2}]\,, \\&& \nonumber
B_{0} = \frac{2\sqrt{3}\lambda_{10}^\frac{3}{2} \lambda_{12}}{\lambda_{24}}\,, \\&& \nonumber
B_{1} = \sqrt{3\lambda_{10}}\lambda_{16} = \sqrt{3\lambda_{10}}\lambda_{21}\,, \\&& \nonumber
B_{2} = 3\lambda_{10} + 3\lambda_{11} + 6\lambda_{10}\lambda_{5}\,.
\end{eqnarray}

\noindent The coefficients are defined as
\begin{eqnarray}
&& \lambda_{11} = \Big(\frac{\lambda_{1}^2}{6\lambda_{1}(2\lambda_{1}-1)} - \frac{1}{6}\Big)\,, \\&&
\lambda_{12} = \lambda_{4} \Big(\frac{3\lambda_{11}^2}{\lambda_{10}} - 6\lambda_{11} + 3\lambda_{10} 
-2 \Big)\,, \\&&
\lambda_{13} = \Big(\lambda_{5} + \frac{\lambda_{11}}{2\lambda_{10}} - \frac{\lambda_{12}}{2\sqrt{3}\lambda_{4}\sqrt{\lambda_{10}}} - \frac{1}{2} \Big)\,, \\&&
\lambda_{14} = \lambda_{15} = \lambda_{17} = \lambda_{18} = (\lambda_{5} - \lambda_{13})\,, \\&&
\lambda_{16} = \lambda_{21} = \frac{\lambda_{12}}{\lambda_{4}}\,, \\&&
\lambda_{19} = \lambda_{20} = \lambda_{22} = \lambda_{23} = \Big(1 + \frac{\lambda_{12}}{\lambda_{4}\sqrt{3\lambda_{10}}} + \lambda_{13} - \lambda_{5} - \frac{\lambda_{11}}{\lambda_{10}} \Big)\,, \\&&
\lambda_{24} = \Bigg[ \frac{\lambda_{12}\lambda_{10}}{\lambda_{4}} \Big(3\lambda_{10} - 3\lambda_{11} - 6\lambda_{5}\lambda_{10} + \sqrt{3\lambda_{10}}\frac{\lambda_{12}}{\lambda_{4}}\Big)\nonumber\\&&
\Big(- 3\lambda_{10} + 3\lambda_{11} + 6\lambda_{5}\lambda_{10} + \sqrt{3\lambda_{10}}\frac{\lambda_{12}}{\lambda_{4}}\Big)\Bigg]\,, \\&&
\lambda_{25} = \Big(1 + \lambda_{13} - \lambda_{5} - \frac{\lambda_{11}}{\lambda_{10}}\Big), \\&& \lambda_{26} = (\lambda_{5} -\lambda_{13})\,.
\end{eqnarray}
where $C_{1}$ and $C_{2}$ are constants of integration. The other parameters are sensitive over the self-interaction potential, i.e., the exponent $n$ in the $\frac{dV}{d\phi}$ term. A particular example of reconstructed $f(R)$ model would be for an inverse power law potential, for example, $n = -4$ gives after solving Eq. (\ref{Mequation1})

\begin{equation}\label{example1}
f(R) \sim C_{1} R^{\frac{3}{10}} + C_{2} R + \frac{5}{22} R^{\frac{3}{5}}\,.
\end{equation}

For different choice of the self-interaction potential, the solution for $f(R)$ is effectively in a combination of power law function of $R$. In order to have GR as a special case from this particular class of models, one of the powers must be unity which is satisfied in the particular example presented in Eq. (\ref{example1}). We note here that the scale factor for a simple power law self-interaction given by Eq. (\ref{exactscale1}), seems to describe a constant deceleration parameter model. For this kind of model, the universe is ever accelerating. No signature flip from a phase of deceleration to acceleration is possible. Since recent observations seems to confirm that there is an evolution of the deceleration parameter with redshift, these classes of models become redundant as far as models of accelerating cosmology is concerned. We present the brief discussion on reconstruction for the sake of a simple illustration.

\section{Combination of Power-Law terms as a potential}
In this section we focus on quadratic potentials for which $\frac{dV(\phi)}{d\phi} \sim \phi$. This is restricted on the ground of validity of the theorem ($n\notin \left\{-3,-1,0,1\right\} $). However, we mean to investigate the scope of this method for interaction potentials of the form $\sim \phi^2 + \phi^d$ which is not barred from the scope afterall. Therefore taking
\begin{equation}
V(\phi) = \frac{1}{2}{\phi}^{2} + \frac{1}{n+1}\phi^{n+1}\,,
\end{equation}
the scalar field evolution can be written as
\begin{equation}
\ddot{\phi} + 3\frac{\dot{a}}{a}\dot{\phi} + \phi + \phi^n = 0\,.
\end{equation}

\begin{figure}[h]
\begin{center}
\includegraphics[width=0.40\textwidth]{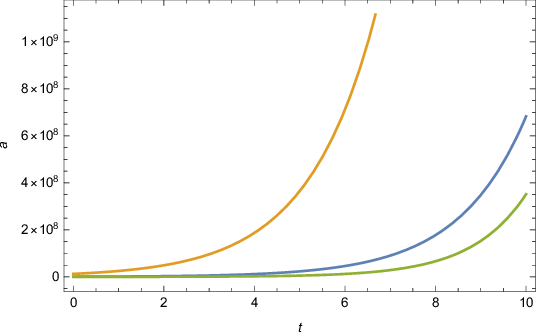}
\caption{Scale factor as a function of cosmic time. Different colors signify different choices of $n$ : $n$ -- Yellow: $n = 3$, Blue: $n = 1.5$, Green: $n = -0.5$.}
\label{mixed_pot_scal}
\end{center}
\end{figure}

\begin{figure}[h]
\begin{center}
\includegraphics[width=0.35\textwidth]{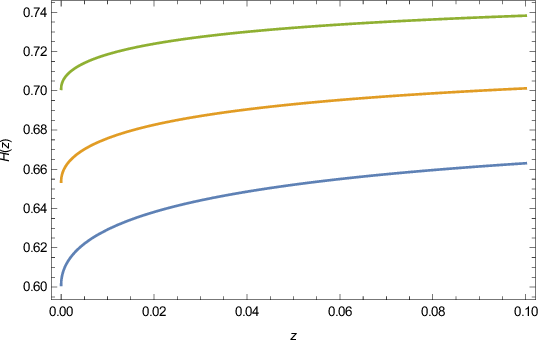}
\includegraphics[width=0.35\textwidth]{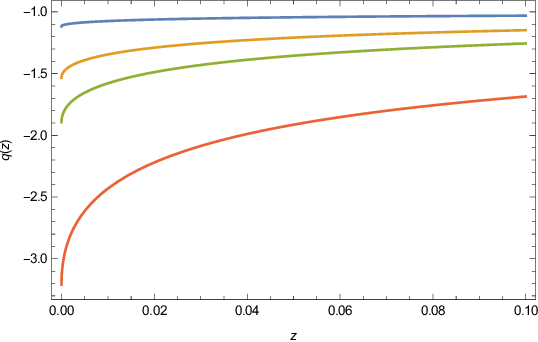}
\caption{$1.$ Left Graph : Hubble parameter as a function of redshift $z$. Different colors signify different choices of $n$ : Blue: $n = 2.5$, Yellow: $n = 3.5$, Green: $n = 4.5$. $2.$ Right Graph : Deceleration parameter as a function of redshift $z$. Different colors signify different choices of $n$ : Red $n = -0.5$, Green $n = 1.5$, Yellow $n = 3$, Blue $n = 5$.}
\label{power_law_hubb}
\end{center}
\end{figure}

On comparison with the general oscillator equation as in Eq.(\ref{gen}), one has $f_{1} = 3\frac{\dot{a}}{a}$, $f_{2} = 1$ and $f_{3} = 1$. The integrability criterion therefore can be used similarly to write an exact form of the scale factor. The point transformed scalar field equation gives the evolution of the scalar field as well. The evolution equation for $a(t)$ can therefore be written as 
\begin{equation}
\frac{6}{(n+3)}\frac{\ddot{a}}{a} + \frac{12n}{(n+3)^2}\frac{\dot{a}^2}{a^2} - 1 = 0\,.
\end{equation}

\noindent The first integral follows as
\begin{equation}
\dot{a}^2 = \frac{(n+3)^2}{18(n+1)} a^2 + \lambda_{0} a^{-\frac{4n}{(n+3)}}\,.
\end{equation}

\noindent Integration of the above equation (using Mathematica) gives the scale factor as
\begin{equation}
\label{exactscale2}
a(t) = \Bigg[\delta_{0} \cosh\Big(\sqrt{\frac{(1+n)}{2}}(t + 6 (3 + n) \delta_{1}\Big)\Bigg]^{\frac{(n+3)}{3(n+1)}}\,,
\end{equation}
where $\delta_0,\, \delta_1$ are constants of integration. We plot the evolution of the scale factor as a function of time in Fig. \ref{mixed_pot_scal} and demonstrate that for some values of $n$, one indeed gets a late-time accelerated expansion. \\

The exact form of the scalar field can be written by solving the point-transformed anharmonic oscillator equation giving
\begin{equation}
\phi(t) = -\frac{2n\sqrt{1- y(t)} y(t)}{3\sqrt{(1+n)}z(t)}\,,
\end{equation}
where
\begin{equation}
y(t) = \cosh\Big(\sqrt{2(1+n)}(t + 6 (3 + n) \delta_{1}\Big)\,,
\end{equation}
and
\begin{eqnarray}\nonumber
&&z(t) =\\&&
{_2}F{_1}\Big[\frac{1}{2},\frac{n}{3(1+n)};\frac{(3+4n)}{(3+3n)};\Big[\cosh\Big(\sqrt{\frac{(1+n)}{2}}(t + 6 (3 + n) \delta_{1}\Big)\Big]^2\Big]\,.
\end{eqnarray}

The Hubble and deceleration parameters as a function of redshift are shown in Fig. \ref{power_law_hubb}. Depending on the choice of $n$ they can provide the correct cosmological dynamics, which may involve the fitting of model parameters against recent data which will be reported in a separate work. As described in the previous section, the $f(R)$ Lagrangian is found by taking the modified Friedmann equation (\ref{fee}). However, the general differential equation that results from the above solution turns out to be extremely non-linear, making the solution for general $n$ a very non-trivial one to write in a closed form. Rather, we focus on a particular case of mixed power-law form of potential which has gained a lot of interest in gravitational physics over the years.

\subsection{Higgs Potential}

Here we study a special combination of power law potential, namely, the Higgs potential, given by
\begin{equation}\label{higgs}
V(\phi) = V_{0} + \frac{1}{2} M^2 {\phi}^{2} + \frac{\lambda}{4}\phi^{4}\,,
\end{equation}
such that $\frac{dV}{d\phi} = M^2 \phi + \lambda \phi^3$. Thus, the scalar field evolution becomes
\begin{equation}
\ddot{\phi} + 3\frac{\dot{a}}{a}\dot{\phi} + M^2 \phi + \lambda \phi^3 = 0\,.
\end{equation}

\begin{figure}[h]
\begin{center}
\includegraphics[width=0.40\textwidth]{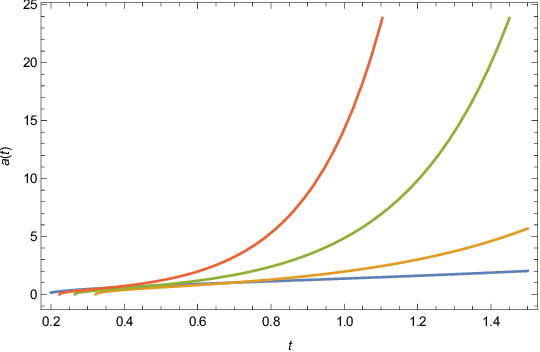}
\includegraphics[width=0.40\textwidth]{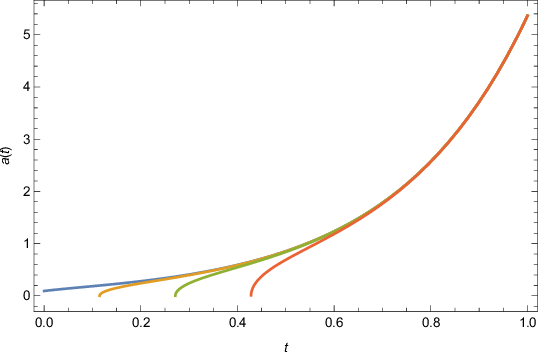}
\caption{$(a)$ Left: Scale factor as a function of time, the mass parameter being varied, $\lambda$ kept fixed at $0.85$. Different colors signify different value of the mass parameter : Blue: $M=1$, Yellow: $M=3$, Green: $M=5$, and Red: $M=7$. $(b)$ Right: Scale factor as a function of time, $\lambda$ being varied and the mass term being kept fixed $M = 3$. Different colors signify different choices of $\lambda$ : Blue: $\lambda=-20$, Yellow: $\lambda=-10$, Green: $\lambda=-0.1$, Red: $\lambda=0.1$.}
\label{higg_scale_fac}
\end{center}
\end{figure}

Comparing with Eq.(\ref{gen}), one has $f_{1} = 3\frac{\dot{a}}{a}$, $f_{2} = M^2$ and $f_{3} = \lambda$. Therefore, the evolution of the scale factor can be written from the theorem of integrability using Eq. (\ref{int-gen}) as
\begin{equation}
\frac{6 \ddot{a}}{(n+3) a} + \frac{12n \dot{a}^{2}}{(n + 3)^{2} a^{2}} = M^{2}\,.
\end{equation}

Transforming the derivatives with respect to time into derivatives with respect to the scalae factor, and integrating twice, one appears at
\begin{equation}\label{exactscale4}
a(t) = \Bigg[\frac{1}{2 M^2} e^{\sqrt{2}Mt} - \lambda e^{-\sqrt{2}Mt}\Bigg]^{\frac{1}{2}}\,.
\end{equation}
We plot the scale factor as a function of time for different choices of $M$ (with $\lambda$ fixed) and different $\lambda$ (with $M$ fixed) in Fig. \ref{higg_scale_fac}. The plots depict an accelerated expansion at late-times. The rate of expansion depends on the choice of parameters as can be seen from the plots. However, the behavior at early times may differ for different choices of $\lambda$ which is also evident from the exact solution in Eq.(\ref{exactscale4}). \medskip

\begin{figure}[h]
\begin{center}
\includegraphics[width=0.40\textwidth]{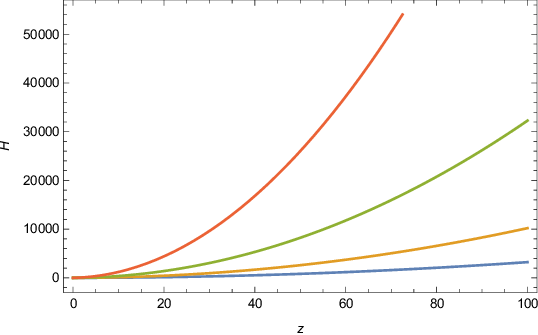}
\includegraphics[width=0.40\textwidth]{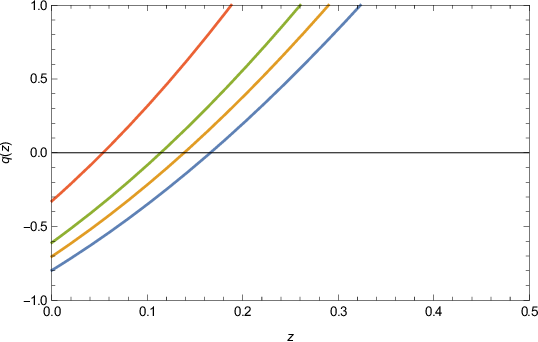}
\includegraphics[width=0.40\textwidth]{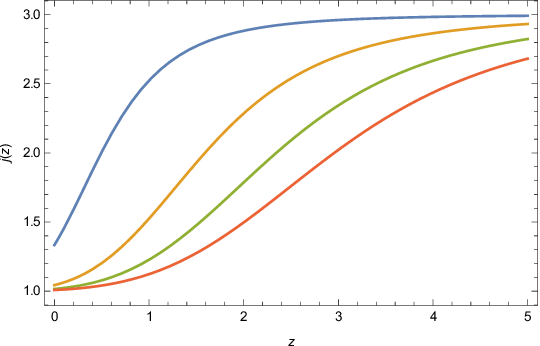}
\caption{$(a)$ Top Left: Hubble parameter as a function of redshift $z$. Different colors signify different values of $M$, which are similar as in Fig.(\ref{higg_scale_fac}). $(b)$ Top Right: Deceleration parameter as a function of redshift $z$. $(c)$ Bottom: Jerk parameter as a function of redshift $z$. Different colors signify different values of $M$, which are similar as in Fig.(\ref{higg_scale_fac}).}
\label{higgs_hubble}
\end{center}
\end{figure}

The scalar field as a function of time is determined from the point-transformed anharmonic oscillator equation which results in
\begin{equation}
\phi\left( t \right) =\phi_{0}\left[ C^{-1} \int \frac{1}{a(t)} dt -T_{0}\right] ^{-1} \frac{1}{a(t)}\,,
\end{equation}
which is representative for the choice of $n=3$ and $\lambda=1$. \medskip

Using the expression of the scale factor from Eq.(\ref{exactscale4}), the scalar field can be written as
\begin{equation}\label{higgsscalar}
\phi(t) = D_{0} \frac{\sqrt{2} M}{\sqrt{\Big[4-\frac{2}{\lambda M^2}e^{2\sqrt{2}Mt}\Big]} {_2}F{_1}\Big[\frac{1}{4},\frac{1}{2};\frac{5}{4};\frac{e^{2\sqrt{2}Mt}}{2\lambda M^2}\Big]}\,,
\end{equation}
where $D_{0}$ is a constant consisting of $\lambda$, $C$ and $\phi_{0}$. \medskip

The Hubble parameter $H(z)$, the deceleration parameter $q(z)$ and the jerk parameter $j(z)$ are plotted as a function of redshift in Fig. \ref{higgs_hubble}. The deceleration parameter $q(z)$ and the jerk parameter $j(z)$ describe a correct description of the present deceleration parameter for some choices of $M$. We expect that from a more detailed data analysis and parameter estimation, one could constrain the acceptable values of $M$.

\subsection{Reconstruction of $f(R)$}
The solutions described in the last section offers the opportunity to reconstruct explicitly the functional form of $f(R)$ that allows such a dynamics for a Higgs potential. The standard method once again will be to use the modified Friedmann Eq. (\ref{fee}) and the solutions in Eq. (\ref{exactscale4}) and Eq. (\ref{higgsscalar}) to deduce the functional form of $f(R)$. However, treating the Friedmann equation in Eq.(\ref{fee}) requires expressing the scalar field and the interaction potential as a function of the Ricci scalar. This is not a straightforward task in the case of a Higgs potential since the scalar field is given by a hypergeometric function and simplifying the function without any further assumptions is very difficult. We study the equation under a late-time approximation, i.e., when $t$ is very large. \\

At late-times a term containing $e^{\alpha t}$ will definitely dominate over a term of $e^{-\alpha t}$ and therefore the scale factor can be approximated as 

\begin{equation}
a(t) \sim \frac{1}{\sqrt{2}M}e^{\frac{M t}{\sqrt{2}}}\,.
\end{equation}

\noindent A similar approximation sees the scalar field $\phi(t)$ behave as

\begin{equation}
\phi(t) \sim D_{1} e^{-\beta t},
\end{equation}

where $D_{1} = \frac{2 D_{0}}{\lambda_{3} \lambda_{1}} \Big(\frac{1}{2\lambda_{3}^2 M^2} \Big)^{\frac{1}{4}}$. $\lambda_{1} = \frac{\Gamma(\frac{5}{4})}{\Gamma(1)}$, $\lambda_{3} = \frac{\Gamma(-\frac{1}{2})}{\Gamma(\frac{3}{4})}$ and $\beta = \sqrt{2}-\frac{1}{\sqrt{2}}$. Since $\beta > 0$, the scalar field clearly decays into a very small value at late-times. Similar behavior is expected from $\dot{\phi}^2$, $\phi^2$ and $\phi^4$ as well. The modified Friedmann equation in Eq.(\ref{fee}) therefore simplifies to
\begin{equation}
\Big(\frac{M^2}{2} - \frac{R}{6}\Big)\frac{df(R)}{dR} + \frac{1}{6}f(R) - \frac{V_{0}}{3} = 0\,.
\end{equation} 

\noindent This can be solved in a straightforward manner to write the $f(R)$ functional as
\begin{equation}\label{solution1}
f(R) = \frac{(C_{1} + 6 R V_{0})}{3 M^2 - R},
\end{equation}
where $C_{1}$ is a constant of integration.

One can also study the system of equations with an early time approximation, i.e., when $t \rightarrow 0$. Under such an approximation, the exponential functions can be written as series expansions, with higher order terms contributing negligibly. Thereafter the scale factor takes a linear form as
\begin{equation}
a(t) = \Big(\frac{1}{2M^2}-\lambda \Big) + \sqrt{2}M t \Big(\lambda + \frac{1}{2M^2} \Big) = b + dt\,,
\end{equation}
and the scalar field takes the form as
\begin{equation}
\phi(t) = \delta_{0} \Bigg(-b - \frac{\sqrt{2}}{M^2}t \Bigg)^{-\frac{1}{2}} \Bigg(\frac{d}{\sqrt{2}M} + \frac{\sqrt{2}}{M^2} t \Bigg)^{-1}\,,
\end{equation}
where $\delta_{0} = \frac{10}{\sqrt{2}} \lambda^{\frac{3}{2}} M D_0$\,. 

The modified Friedmann equation can be written under this approximation as
\begin{eqnarray}\label{solution2}
\frac{d^{2} f(R)}{dR^2} + \frac{(\alpha_{1}R + \beta_{1})}{(\alpha_{2}R + \beta_{2})}\frac{df(R)}{dR} + \frac{(\alpha_{3}R + \beta_{3})}{(\alpha_{4}R + \beta_{4})} f(R) - V_{0} \frac{(\alpha_{5}R + \beta_{5})}{(\alpha_{6}R + \beta_{6})} = 0\,.
\end{eqnarray}

The $\alpha_{i}-s$ and $\beta_{i}-s$ are constant coefficients consisting of $M$, $\lambda_0$, $D_0$ and $V_0$ which we write in this manner for the sake of brevity. A general solution of this equation could not be written in closed form analytically for all ranges of $R$. However, it is easy to note that in the limits $R \rightarrow 0$ and $R \rightarrow \infty$, Eq.(\ref{solution2}) can be written as

\begin{equation}\label{solution3}
\frac{d^{2} f(R)}{dR^2} + \delta_{1}\frac{df(R)}{dR} + \delta_{2} f(R) - \delta_{3} = 0\,.
\end{equation}

$\delta_{i}-$s have different values in the limits $R \rightarrow 0$ and $R \rightarrow \infty$. In the limit $R \rightarrow 0$, $\delta_{1} = \frac{\beta_1}{\beta_2}$, $\delta_{2} = \frac{\beta_3}{\beta_4}$ and $\delta_{3} = V_{0}\frac{\beta_5}{\beta_6}$. In the limit $R \rightarrow \infty$, $\delta_{1} = \frac{\alpha_1}{\alpha_2}$, $\delta_{2} = \frac{\alpha_3}{\alpha_4}$ and $\delta_{3} = V_{0}\frac{\alpha_5}{\alpha_6}$. Solution for $f(R)$ from Eq.(\ref{solution3}) can be written as

\begin{eqnarray}\label{solutionfinal}
f(R) = \frac{\delta_3}{\delta_2} + C_{1} e^{\frac{1}{2} \Big(-\delta_{1} - \sqrt{\delta_{1}^{2} - 4\delta_{2}} \Big) R} + C_{2} e^{\frac{1}{2} \Big(-\delta_{1} + \sqrt{\delta_{1}^{2} - 4\delta_{2}} \Big) R}\,.
\end{eqnarray}

\noindent For the above expression to produce a real solution for $f(R)$, $\delta_{1}^{2} > 4\delta_{2}$\,.

\section{Exponential Potential}
In this section, we extend our study to a case where the self-interaction potential is an exponential function of the scalar field. This case does not exactly fall within the regime of the `anharmonic oscillator equation' treatment, however, we include a very simple example of reconstruction for this case, keeping in mind the huge importance of this choice from a cosmological purview. These potentials are found relevant in higher-order gravity theories, discussed by Whitt \cite{Whitt}, in nonperturbative setups relevant for gaugino condensation as discussed by Carlos, Casas, and Munoz \cite{Carlos}. Moreover, an exponential potential can produce a power law inflation as discussed by Halliwell \cite{Halliwell}. \\

\noindent The scalar field evolution equation for this case becomes
\begin{equation}
\ddot{\phi} + 3\frac{\dot{a}}{a}\dot{\phi} + V_{0} e^{\alpha \phi} = 0\,,
\end{equation}
where $V=(V_0/\alpha)\, e^{\alpha \phi}$\,. \medskip

\noindent We present a very simple set of solution of the above equation as
\begin{equation}\label{exactscale5}
a(t) = (t - t_{0})^{m^2}\,,
\end{equation}
and 
\begin{equation}\label{expophi}
\phi(t) = -\frac{2}{\alpha} \ln(t - t_{0})\,.
\end{equation}

From a quick consistency check, we enforce a restriction on $m^2$ as $m^2 = \frac{(2 + \alpha V_{0})}{6}$.
\begin{figure}[h]
\begin{center}
\includegraphics[width=0.35\textwidth]{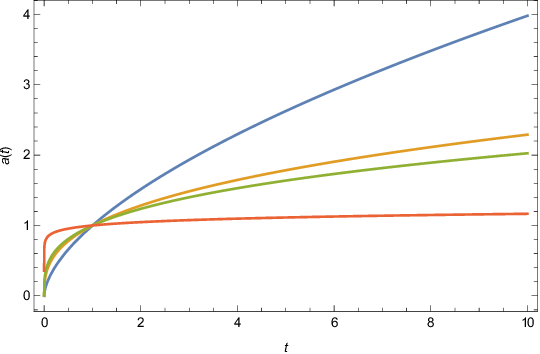}
\caption{Scale factor as a function of time. Different colors indicate different values of $V_0$ : Red: $V_{0} = 10$, Blue: $V_{0} = 1$, Yellow: $V_{0} = - 1$, and Green: $V_{0} = - 10$. The parameter $\alpha$ is fixed at $\alpha = \frac{1}{6}$.}
\label{exp_scale_factor_V}
\end{center}
\end{figure}

\begin{figure}[h]
\begin{center}
\includegraphics[width=0.45\textwidth]{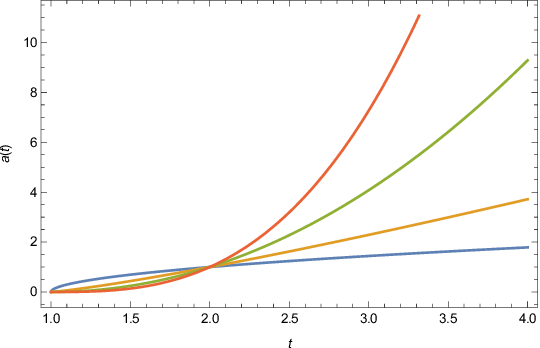}
\caption{Scale factor as a function of time. Different colors indicate different values of $\alpha$, Red: $\alpha = 15$, Blue: $\alpha = 10$, Yellow: $\alpha = 5$, and Green: $\alpha = 1$. The parameter $V_{0}$ is fixed at $V_{0} = 1$.}
\label{exp_scale_factor_alpha}
\end{center}
\end{figure}

In Fig. \ref{exp_scale_factor_V} the scale factor $a(t)$ is plotted for different choices of the parameter $V_0$ for a particular choice of $\alpha = \frac{1}{6}$. For a choice of $V_0 = 10$ one finds a steady expansion with time $t$. However, for other values of $V_0$, the curves eventually show a deceleration. We also plot the scale factor for different choices of the parameter $\alpha$ in Fig. \ref{exp_scale_factor_alpha}. For some values of $\alpha$ the scale factor shows a steep expansion as is evident from the figure.

\begin{figure}[h]
\begin{center}
\includegraphics[width=0.35\textwidth]{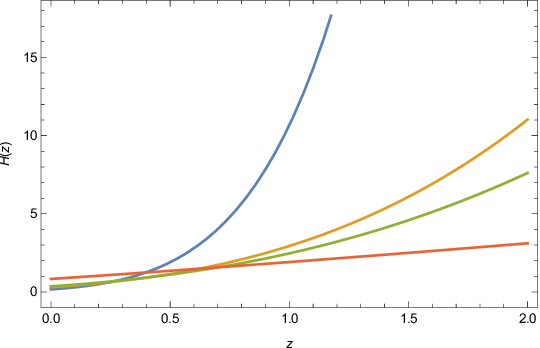}
\caption{Hubble parameter as a function of redshift $z$, $\alpha$ being varied as Red: $\alpha = 3$, Green: $\alpha = 0.16$, Yellow: $\alpha = -0.16$, Blue: $\alpha = -1$.}
\label{exp_hubble}
\end{center}
\end{figure}

\begin{figure}[h]
\begin{center}
\includegraphics[width=0.35\textwidth]{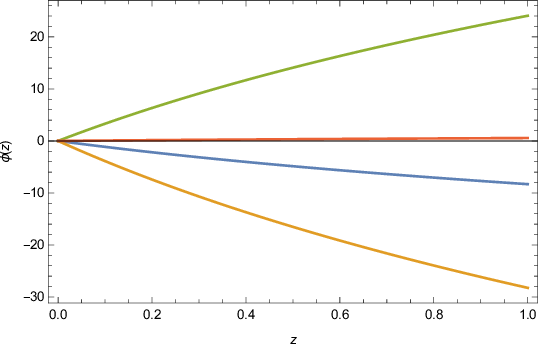}
\caption{Scalar field as a function of redshift, $\alpha$ being varied as Red: $\alpha = 3$, Green: $\alpha = 0.16$, Yellow: $\alpha = -0.16$, Blue: $\alpha = -1$.}
\label{exp_scalar}
\end{center}
\end{figure}

The behavior of Hubble parameter $H$ and the scalar field are studied graphically as a function of redshift $z$ in Figs. \ref{exp_hubble} and \ref{exp_scalar}, respectively.

\subsection{Reconstruction of $f(R)$}
In this section we reconstruct and write explicitly the functional form of $f(R)$ that allows a dynamics as described in the previous section, for an exponential interaction of the scalar field. We use the modified Friedmann equation in Eq.(\ref{fee}) and the solutions in Eq.(\ref{exactscale5}) and Eq.(\ref{expophi}) to deduce the functional form of $f(R)$. Writing the modified Friedmann equation as a function of the Ricci scalar $R$ involves transforming the scalar field and it's first derivative, the potential and the Hubble parameter, all into a function of $R$. The transformed equation for reconstruction looks like

\begin{eqnarray}\nonumber
&&\frac{d^{2} f(R)}{dR^2} - \frac{m_{1}^2}{2m^2}\Bigg(\frac{m^4}{m_{1}^2} - 6\Bigg)\frac{1}{R}\frac{df(R)}{dR} + \frac{1}{6 \alpha m^2}\Bigg(V_{0} + \frac{2}{\alpha}\Bigg)\frac{1}{R}\\&&
 - \frac{m_{1}^2}{12m^2}\frac{f(R)}{R^2} = 0\,.
\end{eqnarray}

Here $m_{1}^2 = m^{2}(m^{2} - 1)$. To solve for a real $f(R)$, $m_{1}^2 < 0$, therefore, $0 < m < 1$. Let us write $m_{1}^2 = - n_{1}^2$ and rewrite the equation as

\begin{eqnarray}\nonumber
&&\frac{d^{2} f(R)}{dR^2} - \frac{n_{1}^2}{2m^2}\Bigg(\frac{m^4}{m_{1}^2} + 6\Bigg)\frac{1}{R}\frac{df(R)}{dR} + \frac{1}{6 \alpha m^2}\Bigg(V_{0} + \frac{2}{\alpha}\Bigg)\frac{1}{R}\\&&
 + \frac{n_{1}^2}{12m^2}\frac{f(R)}{R^2} = 0\,.
\end{eqnarray} 

\noindent The general solution of this equation can be written as
\begin{equation}\label{solutionexpo}
f(R) = \frac{f_{1}(R)}{f_{2}} + C_{1} R^{G_1} + C_{2} R^{G_2}\,.
\end{equation}

\noindent The functions and coefficients $f_{1}(R)$, $f_{2}$, $G_1$ and $G_2$ are defined below.
\begin{eqnarray}\nonumber
&& f_{1}(R) = \Bigg[2B R^{a_1} \Bigg(-R^{a_2} + A R^{a_3} + \sqrt{(1 + 2A + A^2 - 4G)} R^{a_4} \\&&
 + (1 - A) R^{a_5} + \sqrt{(1 + 2A + A^2 - 4G)} R^{a_6} \Bigg)\Bigg]\,,
\end{eqnarray}
where
\begin{equation}
a_1 = \frac{1}{2}-\frac{A}{2}-\frac{\sqrt{(1 + 2A + A^2 - 4G)}}{2}\,,
\end{equation}

\begin{eqnarray}\nonumber
&&a_2 = \sqrt{(1 + 2A + A^2 - 4G)} + \frac{\sqrt{G}}{2} \Big(\frac{(1-A)}{\sqrt{G}}\\&&
 - \frac{\sqrt{(1+2A+ A^2 -4G)}}{\sqrt{G}}\Big)\,,
\end{eqnarray}

\begin{eqnarray}\nonumber
&&a_3 = \sqrt{(1 + 2A + A^2 - 4G)} + \frac{\sqrt{G}}{2} \Big(-\frac{(1-A)}{\sqrt{G}}\\&&
 - \frac{\sqrt{(1+2A+ A^2 -4G)}}{\sqrt{G}}\Big)\,,
\end{eqnarray}

\begin{eqnarray}\nonumber
&&a_4 = \sqrt{(1 + 2A + A^2 - 4G)} + \frac{\sqrt{G}}{2} \Big(-\frac{(1-A)}{\sqrt{G}}\\&&
 - \frac{\sqrt{(1+2A+ A^2 -4G)}}{\sqrt{G}}\Big)\,,
\end{eqnarray}

\begin{equation}
a_5 = \frac{\sqrt{G}}{2} \Big(-\frac{(1-A)}{\sqrt{G}} + \frac{\sqrt{(1+2A+ A^2 -4G)}}{\sqrt{G}}\Big)\,,
\end{equation}

\noindent and

\begin{equation}
a_6 = \frac{\sqrt{G}}{2} \Big(-\frac{(1-A)}{\sqrt{G}} + \frac{\sqrt{(1+2A+ A^2 -4G)}}{\sqrt{G}}\Big)\,,
\end{equation}

\begin{eqnarray}\nonumber
&&f_2 = \Bigg[ (1-A+\sqrt{1+2A+A^2 -4G})(-1+A\\&&
 + \sqrt{1+2A+A^2 -4G})\sqrt{1+2A+A^2 -4G}\Bigg]\,,
\end{eqnarray}

\begin{equation}
G_1 = \frac{\sqrt{G}}{2} \Bigg(\frac{(1+A)}{\sqrt{G}} - \frac{\sqrt{(1+2A+ A^2 -4G)}}{\sqrt{G}}\Bigg)\,,
\end{equation}
and
\begin{equation}
G_2 = \frac{\sqrt{G}}{2} \Bigg(\frac{(1+A)}{\sqrt{G}} + \frac{\sqrt{(1+2A+ A^2 -4G)}}{\sqrt{G}}\Bigg)\,.
\end{equation}

\noindent The coefficients $A$, $B$ and $G$ are defined as
\begin{equation}
A = \frac{n_{1}^2}{2m^2}\Bigg(6 + \frac{m^4}{n_{1}^2}\Bigg)\,,
\end{equation}

\begin{equation}
B = \frac{1}{6 \alpha m^2} \Bigg( V_{0} + \frac{2}{\alpha}\Bigg)\,,
\end{equation}
and
\begin{equation}
G = \frac{n_{1}^2}{12m^2}.
\end{equation}

We simplify Eq.(\ref{solutionexpo}) and demonstrate one particular case, for instance for $m = \frac{1}{2}$. After some straightforward manipulation, one can calculate all the coefficients, and write $f(R)$ as

\begin{equation}\label{fRexpo}
f(R) \sim (C_{1} + 64) R^{-\frac{1}{16}} + C_{2} R^{\frac{55}{16}} - 8 R^{-\frac{23}{16}}\,.
\end{equation}

\section{Viability of $f(R)$ models}
In this section we briefly review the existing works on the viability issues of $f(R)$ theories and analyze our models comparatively. Due to their simplicity, $f(R)$ models are perhaps the most popular and well-studied alternative theory of gravity. However, there are certain criteria for the viability that has to be accounted for, for instance, it is expected that an $f(R)$ model must satisfy cosmological requirements such as smooth transitions between cosmological eras. In some cases, the transition from radiation era can be problematic to describe as discussed by Faraoni \cite{fara1}, Amendola, Polarski and Tsujikawa \cite{amen1}, Amendola, Gannouji, Polarski and Tsujikawa \cite{amen2}. Certains $f(R)$ models pose non-linear instability which makes constructing relativistic stars extremely non-trivial. These instabilities were discussed by Dolgov and Kawasaki \cite{dol} for a typical model like $f(R) \sim R - \frac{\mu^4}{R}$. For further discussions and possible avenues of avoiding the instability one may refer to the works of Nojiri and Odintsov \cite{nojiodi1}, Baghram, Farhang, and Rahvar \cite{bagh}, Faraoni \cite{fara1,fara2}, Cognola and Zerbini \cite{cogno}. In general, $f(R)$ models are stable under the necessary and sufficient condition that $\frac{d^{2}f(R)}{dR^2} \geq 0$ and unstable if $\frac{d^{2}f}{dR^2} < 0$. These give rise to certain restrictions over the parameters of the functional form of $f(R)$, studied in the context of star solutions in $f(R)$ theories by  Seifert \cite{seif}. The stability conditions were re-adressed by Sawicki and Hu \cite{hu} in the context of cosmological perturbations. Generalization of $f(R)$ theories by including higher curvature terms may contain ghost fields, but in general $f (R)$ gravity is does not contain ghost fields (Ghost fields are massive negative norm states who are common when trying to generalize GR).  \\

A complete set of physical criteria to select a particular form of theory capable of matching all the required data for all scales is yet to be found. Different observational aspects of $f(R)$ dark energy models that satisfy cosmological and local gravity constraints fairly well were discussed by Copeland, Sami and Tsujikawa \cite{cope}, Das, Banerjee and Dadhich \cite{das}.  \\

In the present case, the reconstruction technique in almost all of the cases produces $f(R)$ as a combination of power-law terms of Ricci scalar, i.e., 
\begin{equation}\label{fR}
f(R) \sim \Sigma a_{i} R^{b_i} \sim a_{1}R^{b_1} + a_{2}R^{b_2} + a_{3}R^{b_3}... 
\end{equation}

Such examples are known to play an important role in $f(R)$ cosmology. For instance, the model with $f (R) = R + \alpha R^2$ ($\alpha > 0$) falls under the category of Eq.(\ref{fR}) and can lead to the accelerated expansion of the universe because of the presence of the $\alpha R^2$ term. To be specific, this was the very first model proposed to account for inflation by Starobinsky \cite{staro}. Theories carrying the form $f (R) \sim R - \frac{\alpha}{R^n}$ ($\alpha > 0, n > 0$) are in general relevant for a dynamical description of dark energy, studied by Paul, Debnath and Ghoshe \cite{ghoshe}. However, such models can fail to describe standard matter-dominated epoch of cosmology. Nojiri ad Odintsov \cite{nojiodi1} showed that this can be avoided by further adding an $R^2$ or an $ln R$ term to the action. \\

An exponential $f(R)$ model was described by Cognola et al. \cite{cognola} and Linder \cite{linder}. The form of the Lagrangian was taken as
\begin{equation}\label{expofR}
f(R) \sim c_{1} + c_{2} R + c_{3} e^{-d R}\,.
\end{equation}

Bamba, Geng and Lee \cite{bambageng} studied the possibility of realizing a late-time acceleration in this class of theories. The evolution of an effective equation of state and transition from inflation to dark energy phase was studied in details by Bamba et al. \cite{bamba2}. In the present case, the general forms of reconstructed $f(R)$ models can be interpreted as a higher curvature approximation of the exponential case as in Eq.(\ref{expofR}), producing a sum of power functions of Ricci scalar afterall. \medskip

\begin{figure}[h]
\begin{center}
\includegraphics[scale=0.7]{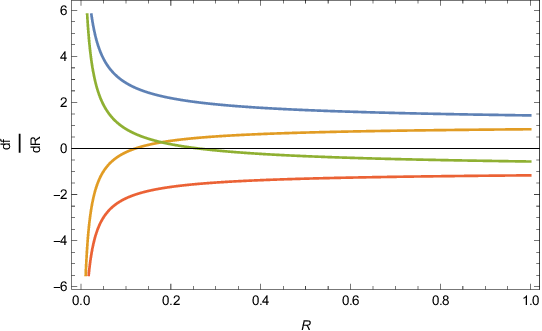}
\includegraphics[scale=0.7]{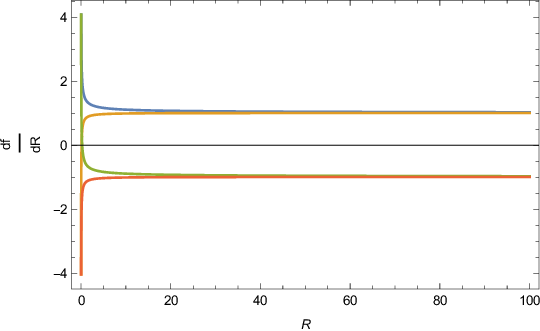}
\caption{Evolution of $\frac{df}{dR}$ as a function of $R$ for $V(\phi) \sim \phi^{(n+1)}$. The graph on the left shows the evolution in low curvature limit whereas the right graph shows the evolution in a high curvature limit. Different color of the curves suggest different choices over the choice of parameters $C_{1}$ and $C_{2}$ : Blue $\rightarrow C_{1}, C_{2} > 0$, Yellow $\rightarrow C_{1} < 0, C_{2} > 0$, Green $\rightarrow C_{1} > 0, C_{2} < 0$ and Red $\rightarrow C_{1}, C_{2} < 0$.}
\label{fprimepowerlaw}
\end{center}
\end{figure}

\begin{figure}[h]
\begin{center}
\includegraphics[scale=0.7]{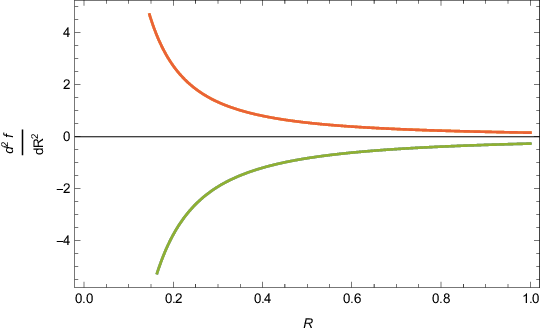}
\includegraphics[scale=0.7]{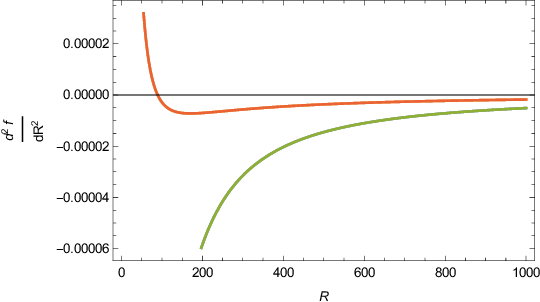}
\caption{Evolution of $\frac{d^{2}f}{dR^2}$ as a function of $R$ for $V(\phi) \sim \phi^{(n+1)}$. The graph on the left shows the evolution in low curvature limit whereas the right graph shows the evolution in a high curvature limit. Different color of the curves suggest different choices over the choice of parameters $C_{1}$ and $C_{2}$ : Red $\rightarrow C_{1}, C_{2} < 0$; $C_{1} < 0, C_{2} > 0$, Green $\rightarrow C_{1}, C_{2} > 0$; $C_{1} > 0, C_{2} < 0$.}
\label{fdprimepowerlaw}
\end{center}
\end{figure}

The detection of gravitational waves (GW) presents relativists a primise to validate the present modifications of gravity and put constraints on the parameters of the theory \cite{abott,nunes}. GW signals from compact black hole binaries indicate their generation and propagation mechanisms and recently, they are used to study possible restrictions over modified gravity \cite{lomb,jana1}. Our interest falls in the cases where constraints on $f(R)$ gravity are studied from $GW170817$, for example by Jana and Mohanty \cite{jana2}. Here, the authors discussed viable $f(R)$ models in the context of a characteristic uncertainty in the observed mass spectrum of the GW observation $GW170817$ and put some restriction on the Lagrangian of the theory. A model independent bound 
\begin{equation}
|F(R_0) - 1| < 3 \times 10^{-3}\,,
\end{equation}
is enforced where $F(R) = \frac{df(R)}{dR}$ and $R_0$ is the present value of the curvature of our universe. In the present case, the $f(R)$ models given by Eqs. (\ref{solution}, \ref{solution1}, \ref{solution2}, \ref{solution3}) all produce a combination of power law terms. The constant coefficients depend on the self-interaction potential of the scalar field under consideration. Using the above restriction, one can put constraints on the choice of the parameters of the theories in a straightforward manner.

\begin{figure}[h]
\begin{center}
\includegraphics[scale=0.7]{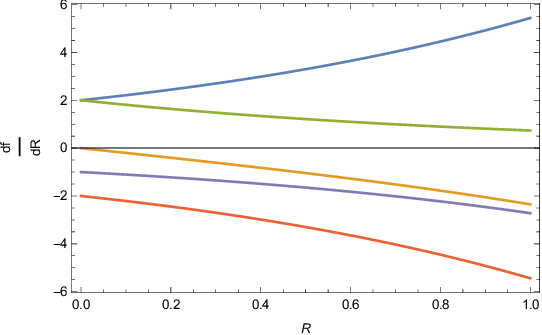}
\includegraphics[scale=0.7]{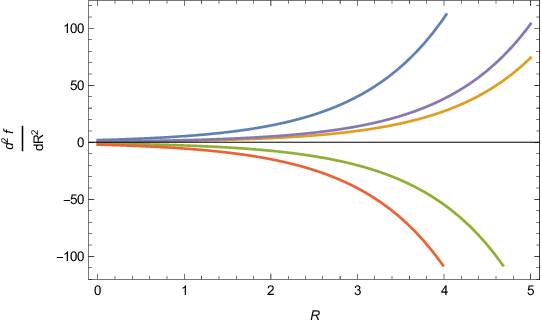}
\caption{Evolution of $\frac{df}{dR}$ and $\frac{d^{2}f}{dR^2}$ as a function of $R$ for Higgs interaction potential with an early-time approximation. The qualitative behavior remains the same for low and high curvature limit. The graph on the left shows the evolution of $\frac{df}{dR}$ whereas the right graph shows the evolution of $\frac{d^{2}f}{dR^2}$. Different color of the curves suggest different choices of the constants $C_{1}$, $C_{2}$, $p_{2}$ and $p_{3}$. (a) For the graph on top, Blue $\rightarrow C_{1}, C_{2}, p_{2}, p_{3} > 0$, Violet $C_{1} = -0.2; C_{2}, p_{2}, p_{3} > 0$, Green $\rightarrow C_{1}, C_{2}, p_{2}, p_{3} < 0$, Yellow $\rightarrow C_{1}, C_{2}, p_{2} < 0; p_{3} > 0$, Red $\rightarrow C_{1}, C_{2} < 0; p_{2}, p_{3} > 0$. (b) For the graph below, Blue $\rightarrow C_{1}, C_{2}, p_{2}, p_{3} > 0$, Violet $C_{1}, p_{2}, p_{3} > 0; C_{2} = -0.5$, Yellow $\rightarrow C_{1} = -0.3; C_{2}, p_{2}, p_{3} > 0$, Green $\rightarrow C_{1}, p_{2}, p_{3} > 0; C_{2} = -2$, Red $\rightarrow C_{1} = -3; C_{2}, p_{2}, p_{3} > 0$.}
\label{fhiggsearly}
\end{center}
\end{figure}

The viability of the discussed solutions are further checked through the positivity of the first and second derivative of $f(R)$ with respect to Ricci curvature ($f'(R)$ and $f''(R)$) (for detailed methodology and discussions we refer to the review of Silvestri and Trodden \cite{silvestri}. Positivity of $\frac{df(R)}{dR}$ ensures the positivity of effective energy density. On the other hand, positivity of $\frac{d^{2}f(R)}{dR^2}$ ensures the stability of the model. 

\begin{figure}[h]
\begin{center}
\includegraphics[scale=0.7]{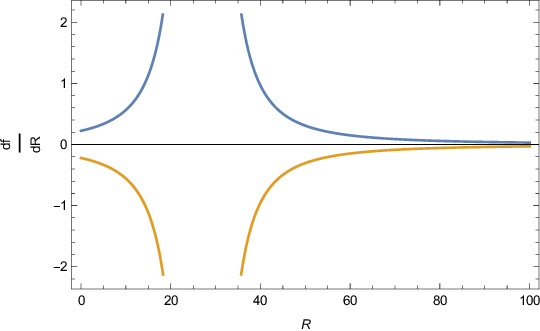}
\includegraphics[scale=0.7]{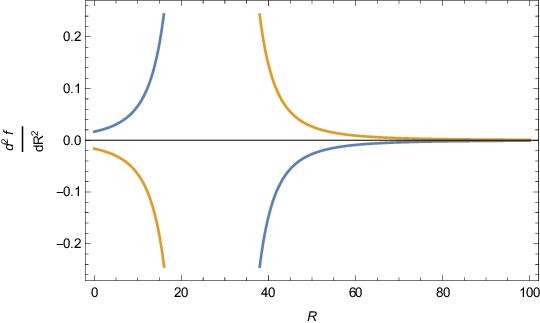}
\caption{Evolution of $\frac{df}{dR}$ and $\frac{d^{2}f}{dR^2}$ as a function of $R$ for Higgs interaction potential with a late-time approximation. The qualitative behavior remains the same for low and high curvature limit. The graph on the left shows the evolution of $\frac{df}{dR}$ whereas the right graph shows the evolution of $\frac{d^{2}f}{dR^2}$. Different color of the curves suggest different choices of the constants $C$ and $V_{0}$, while the choice of the mass parameter is fixed at $M = 3$. In both of the graphs, Blue $\rightarrow C > 0, V_{0} > 0$ and Yellow $\rightarrow C > 0, V_{0} < 0$.}
\label{fhiggslate}
\end{center}
\end{figure}

Fig. \ref{fprimepowerlaw} shows the evolution of $f'(R)$ as a function of $R$, for simple power law self-interaction potential. We show the plots for the specific example presented by Eq. (\ref{example1}). The positivity of $f'(R)$ and $f''(R)$ depends on the signatures of the constants of integration $C_{1}$ and $C_{2}$. The blue curve shows the evolution for $C_{1}, C_{2} > 0$ and the only choice for which the evolution is positive throughout. All the other alternatives show a departure from positivity either at very low curvature or at high curvature. This behavior qualitatively remains the same at high curvature as well. The graphs in Fig. \ref{fdprimepowerlaw} shows the evolution of $f''(R)$ as a function of $R$ in different ranges of the value of Ricci curvature. The red curve shows the evolution for $C_{1}, C_{2} < 0$ and $C_{1} < 0, C_{2} > 0$ while the green curve is for $C_{1}, C_{2} > 0$ and $C_{1} > 0, C_{2} < 0$. For all the possible choices the evolution shows a departure from positivity at a high curvature. \\

\begin{figure}[h]
\begin{center}
\includegraphics[scale=0.7]{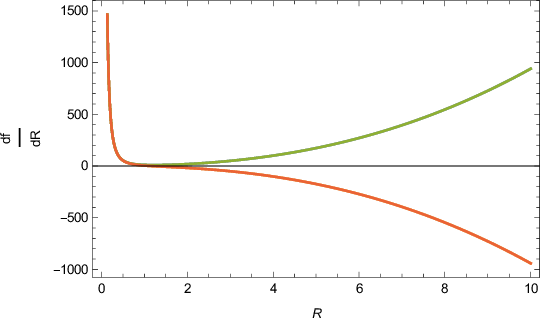}
\includegraphics[scale=0.7]{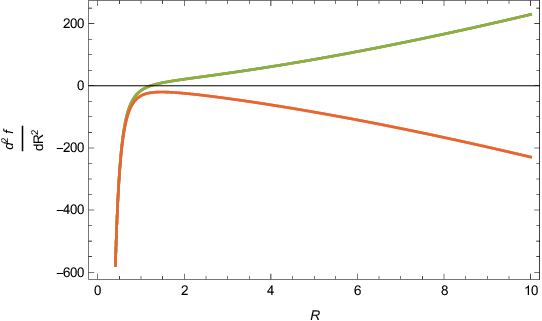}
\caption{Evolution of $\frac{df}{dR}$ and $\frac{d^{2}f}{dR^2}$ as a function of $R$ for an exponential interaction potential. The graph on the top shows the evolution of $\frac{df}{dR}$ whereas the bottom graph shows the evolution of $\frac{d^{2}f}{dR^2}$. Different color of the curves suggest different choices of the constants $C_{1}$ and $C_{2}$. In both of the graphs, Green $\rightarrow C_{1}, C_{2} > 0; C_{1} < 0, C_{2} > 0$ and Red $\rightarrow C_{1} > 0, C_{2} < 0 ; C_{1}, C_{2} < 0$.}
\label{fexpo}
\end{center}
\end{figure}

Fig. \ref{fhiggsearly} shows the evolution of $f'(R)$ and $f''(R)$ as a function of $R$ for Higgs power law interaction potential given by Eq. (\ref{higgs}). Since a general solution for $f(R)$ for all epoch was difficult to come by, we present the graphical study for the early time approximation given by Eq.(\ref{solutionfinal}). Different color of the curves suggest different choices of the constants $C_{1}$, $C_{2}$, $p_{2}$ and $p_{3}$. As can be seen from the figure, the positivity of $f'(R)$ and $f''(R)$ depends heavily on the signature of the parameters $C_{1}$ and $C_{2}$ while for different signatures of the parameters $p_{1}$ and $p_{2}$, the overall qualitative behavior of the plots remain the same. \\

Fig. \ref{fhiggslate} shows the evolution of $f'(R)$ and $f''(R)$ as a function of $R$ for Higgs power law interaction potential under a late-time approximation given by Eq. (\ref{solution1}). Different color of the curves suggest different choices of the parameters $C$ and $V_{0}$. In both of the graphs, the blue curve shows the evolution for $C > 0$, $V_{0} > 0$ and the yellow curve shows the evolution for $C > 0$, $V_{0} < 0$. It is apparent from the graph that the evolution has a singularity at a finite value of $R$ depending on the choice of the value of $M$, at $R = 3M^2$. \\

Fig. \ref{fexpo} shows the evolution of $f'(R)$ and $f''(R)$ as a function of $R$ for an exponential power law interaction potential. The particular case of reconstructed $f(R)$ model given by Eq. (\ref{fRexpo}) is considered. Different color of the curves suggest different choices of the parameters of the models, i.e., $C_{1}$ and $C_{2}$. In both of the graphs, the green curve represents a case where either $C_{1}, C_{2} > 0$ or $C_{1} < 0, C_{2} > 0$. On the other hand, the red curve represents a case where either $C_{1} > 0, C_{2} < 0$ or $C_{1}, C_{2} < 0$.

\section{Conclusions}
$f(R)$ gravity is the most simple modification of GR that provides a direct opportunity to produce late-time accelerated expansion, as well as the early time behavior of the universe. In the present case, we consider $f(R)$ gravity with a minimally coupled self-interacting scalar field in the action. By using some particular forms of the potential such that the resulting set of Friedmann equations become analytically solvable. We study simple power law potentials, combination of power law potentials (general case, and the particular case of quadratic plus quartic case), and an exponential potential, potentials which are extremely well-documented and physically significant for scalar field cosmological scenarios.   \\

We adopt a reverse engineering approach to study the behavior of $f(R)$ Lagrangan that allow simple and viable cosmological dynamics. However, no apriori information regarding the scale factor is assumed to begin with, except the assumption of integrability of the scalar field evolution equation. Used properly, this approach stands true for any physical system that can be written in the form of a classical anharmonic oscillator equation of the form of Eq. (\ref{gen}). Although the work presented here is restricted only for a few potentials only, this clearly carries more applicability. For instance, one may try and generalize the system by adding an additional fluid distribution alongwith the massive scalar field in the system of equations given by Eq. (\ref{SET}), Eq. (\ref{fe1}) and Eq. (\ref{fe2}). In such a case the reconstruction scheme goes as described, extracting the scale factor and scalar field from the klein-gordon equation, thereafter using Eq. (\ref{fe2}) for the reconstruction scheme and Eq. (\ref{fe1}) to determine the density of the additional fluid component. In this regard we would also like to stress upon the fact that a scalar field itself alongwith a suitable choice of potential can mimic the evolution of a dust ball (see for instance \cite{gonca}). \\

However, the solutions of scale factor obtained by means of the theorem are by no means unphysical. While, a simple power law potential produces a simple constant deceleration parameter solution, a Higgs potential generates a more accurate description of the accelerated expansion as is demonstrated by a study of the evolution of deceleration parameter as in Fig. \ref{higgs_hubble}. Moreover, the resulting $f(R)$ Lagrangians as given in Eqs. (\ref{solution}, \ref{solution1}, \ref{solution2}, \ref{solution3}) can be shown to reproduce many existing viable $f(R)$ models, directly or under particular limits. \\

We conclude the manuscript with the note that this is indeed a simple case, and the motivation behind exploiting the anharmonic oscillator theorem is largely mathematical. However, it shows that applied properly to physcial systems such a theorem can help reduce the non-linearity of a system and manifest general applications overall. It would be even more interesting to implement the procedure in other alternatives to GR as well.

\begin{acknowledgements}
SC sincerely thanks Prof. Sayan Kar for insightful discussions. SC was supported by the National Post-Doctoral Fellowship (file number: PDF/2017/000750) from the Science and Engineering Research Board (SERB), Government of India. The work of KB was partially supported by the JSPS KAKENHI Grant Number JP 25800136 and Competitive Research Funds for Fukushima University Faculty
(18RI009).
\end{acknowledgements}

\section*{Conflict of interest}
The authors declare that they have no conflict of interest.

\end{document}